\documentclass[showpacs,amsmath,amssymb,prb,twocolumn,superscriptaddress]{revtex4-1}
\usepackage{graphicx}
\usepackage{dcolumn}
\usepackage{bm}

\begin{document}

\title{Interplay between electron-electron and electron-vibration interactions on the thermoelectric properties of molecular junctions}

\author{C. A. Perroni} 
\affiliation{CNR-SPIN and Dipartimento di Fisica, Universita' degli Studi di Napoli ''Federico II'',\\
Complesso Universitario Monte S. Angelo, Via Cintia, I-80126 Napoli, Italy}
\author{D. Ninno}
\affiliation{CNR-SPIN and Dipartimento di Fisica, Universita' degli Studi di Napoli ''Federico II'',\\
Complesso Universitario Monte S. Angelo, Via Cintia, I-80126 Napoli, Italy}
\affiliation{IMAST S.c.ar.l.-Technological District on Engineering of polymeric and composite Materials and Structures, Piazza Bovio 22, I-80133 Napoli, Italy }
\author{V. Cataudella}
\affiliation{CNR-SPIN and Dipartimento di Fisica, Universita' degli Studi di Napoli ''Federico II'',\\
Complesso Universitario Monte S. Angelo, Via Cintia, I-80126 Napoli, Italy}



\begin {abstract}
The linear thermoelectric properties of molecular junctions are theoretically studied close to room temperature within a model including electron-electron and electron-vibration interactions on the molecule.  A nonequilibrium adiabatic approach is generalized to include large Coulomb repulsion through a self-consistent procedure and applied to the investigation of large molecules, such as fullerenes, within the Coulomb blockade regime. The focus is on the phonon thermal conductance which is quite sensitive to the effects of strong electron-electron interactions within the intermediate electron-vibration coupling regime. The electron-vibration interaction enhances the phonon and electron thermal conductance, and it reduces the charge conductance and the thermopower inducing a decrease of the thermoelectric figure of merit. For realistic values of junction parameters, the peak values of the thermoelectric figure of merit are still of the order of unity since the phonon thermal conductance can be even smaller than the electron counterpart.  
\end {abstract}

\maketitle



\section{Introduction}

The direct conversion of temperature differences to electric voltage and vice versa take place in solid state systems.  
These thermoelectric effects can be strong enough in some semiconducting materials to allow either the fabrication of devices converting wasted heat into electrical energy or the realization of solid-state coolers \cite{nolas,ioffe}. A fundamental parameter to quantify the energy conversion efficiency is the dimensionless figure of merit $ZT=G S^2 T /G_K$, where $G$ is the electrical conductance, $S$ the thermopower, $T$ the absolute temperature, and $G_K=G_K^{el}+G_K^{ph}$ is the total thermal conductance, with $G_K^{el}$ and $G_K^{ph}$ electron and phonon thermal conductance, respectively. Indeed, in order to improve the efficiency, mutually contrasting transport properties of the same material have to be optimized.  For instance, in  metals, $ZT$ is typically limited by the Wiedemann-Franz law \cite{kittel}. Large effort is currently made in material science to get bulk values of $ZT$ larger than $1$ and to use solid state systems for actual thermoelectric devices \cite{nolas,shakouri,biswas}. 

Recently, the possibility of controlling materials at the nanoscale has been exploited to optimize the thermoelectric efficiency. \cite{shakouri,koumoto,sothmann} For example, a maximum $ZT\simeq 2.4$ has been observed at room temperature in a thin-film thermoelectric device \cite{venka}. High values of $ZT$ have been reported in quantum dot superlattices \cite{harman} and in semiconductor nanowires \cite{hochbaum}, where phonon confinement can lead to a lower phonon thermal conductance \cite{dressel,murphy1}. Actually, a significant reduction in lattice thermal conductivity  is considered as the main route for having high $ZT$ in low-dimensional materials \cite{venka1}.  The improvement of thermoelectric efficiency can also derive from the discreteness of energy levels in nanostructures resulting into a violation of the Wiedemann-Franz law \cite{Mahan}.  Finally, in nanoscopic Coulomb-coupled systems, the thermoelectric properties can be optimized by exploiting the Coulomb blockade regime and changing the gate voltage
\cite{sothmann}.  

Molecular devices can be efficient for conversion of heat into electric energy since both phonon and electron properties can contribute to increase the thermoelectric figure of merit $ZT$ \cite{arad,dubi}. Indeed, the emerging field of molecular thermoelectrics  has attracted a lot of attention in recent years \cite{majum1,majum2,finch,murphy,galperin,koch,leijnse}. 
The thermoelectric properties of molecular junctions are also interesting in that they can provide useful informations on charge and energy transport otherwise difficult to obtain, such as  the type of carriers (electros/holes) dominating the transport \cite{majum1,majum2,datta,cuevas,datta1}. Measurements of thermoelectric properties  have been performed in junctions with fullerene ($C_{60}$) \cite{majum2} finding a high value of the molecular thermopower ($S$ of the order of $-30$ $\mu V$/K). In these experiments, three different metallic electrodes (platinum, gold, and silver) have been considered achieving a more controllable alignment between Fermi level and molecular orbitals (whose energy separation is still of the order of $0.5$ eV). However, the application of a gate voltage  remains elusive in these kinds of measurements. Moreover, heat transport in molecular devices remain poorly characterized due to experimental challenges \cite{dubi,wang1,wang2,meier} or limited to a range where transport is elastic \cite{cuevas1} .

In molecular junctions, intramolecular electron-electron and electron-vibration interactions typically constitute the largest energy scales affecting the thermoelectric properties.  \cite{cuevas,galperin1,costi} Moreover, the center of mass oscillation of the molecule \cite{Park}, or thermally induced acoustic phonons \cite{Qin} can be an additional source of coupling between electronic and vibrational degrees of freedom. The effects of intramolecular interactions on the transport properties have been studied in the regime of linear response and fully out-of-equilibrium by different theoretical tools \cite{cuevas,galperin1}. The thermopower $S$ and the thermoelectric figure of merit $ZT$  have been found to be  sensitive to the strength of intramolecular interactions 
\cite{koch,leijnse,galperin,yang,zianni,aharon,liu,ren,tagani,arra}. 
However, the phonon thermal contribution $G_K^{ph}$ to the figure of merit $ZT$ has been calculated only at a perturbative level  of the electron-vibration coupling \cite{hsu}.

In devices with large molecules or carbon nanotube quantum dots \cite{cava1}, a nonequilibrium adiabatic approach has been introduced for spinless electrons exploiting the low energy of the relevant vibrational degrees of freedom \cite{mozy,pistol,hussein,alberto,alberto1}. This method is semiclassical for the vibrational dynamics, but it is valid for arbitrary strength of electron-vibration coupling.
Within this approach, we have recently implemented a self-consistent calculation for electron and phonon thermal conductance focusing on the effects of the electron-vibration coupling \cite{perroni2}. 

In this paper, we have studied the thermoelectric properties of a molecular junction with electron-electron and electron-vibration interactions within the linear response regime focusing on a self-consistent calculation of the phonon thermal conductance $G_K^{ph}$ close to room temperature. The nonequilibrium adiabatic approach is generalized to treat finite strong Coulomb interactions within a junction model which takes into account the interplay between the low frequency center of mass oscillation of the molecule and the electronic degrees of freedom within the Coulomb blockade regime. Parameters appropriate for junctions with $C_{60}$ molecules are considered. We have found that, within the intermediate electron-vibration coupling regime, the effects of electron-electron interactions can enhance  $G_K^{ph}$, which, as a function of the gate voltage,  acquires a behavior similar to that of electron thermal conductance. The electron-vibration interaction induces an increase of the phonon and electron thermal conductance, and, at the same time, a decrease of both the charge conductance and the thermopower. The overall effect is a reduction of the thermoelectric figure of merit. Interestingly, for realistic parameters of the model, the peak values of $ZT$ are still of the order of unity. This effect is ascribed to the magnitude of the phonon thermal conductance which can be smaller than the electronic counterpart in a large range of gate voltages.    

The paper is organized as follows. In Sec. II, the model of molecular junction is proposed. In Sec. III, the adiabatic approach generalized for strong local Coulomb interactions is explained. In Sec. IV, the results within the adiabatic approach are discussed. The paper is closed by Appendix A, where the comparison between different treatments of the large Coulomb repulsion  is made within the Coulomb blockade regime.

\section{Molecular junction model}

In this paper, we analyze the Anderson-Holstein model, which is a reference model for molecular junctions. \cite{cuevas,Haug}  
The molecule is modeled as a single electronic level locally interacting with a single vibrational mode. 
In junctions with $C_{60}$ molecules, attention can be focused on a molecular electronic orbital which is sufficiently separated in energy from other orbitals. \cite{perroni2,Lu,natelson,mra} In this paper, we will consider model parameters appropriate to $C_{60}$ molecular junctions. 

\begin{figure}
\centering
\includegraphics[width=8.5cm,height=4cm]{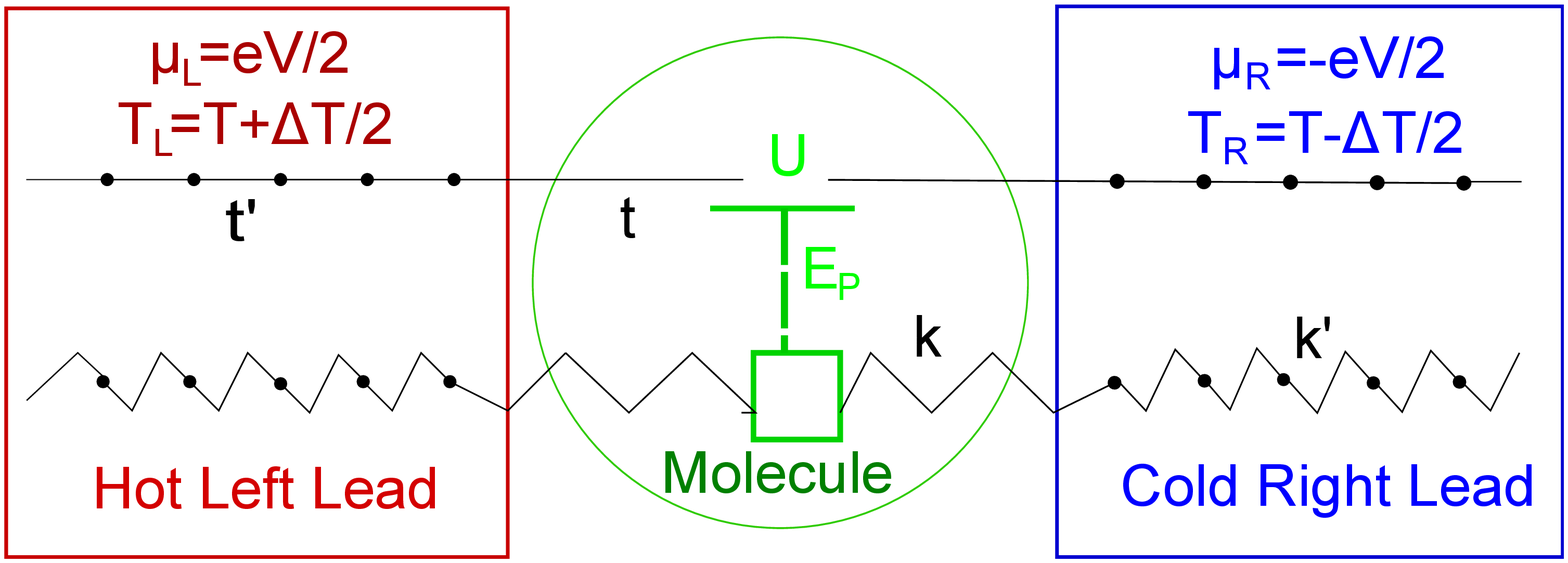}
\caption{(Color online) Sketch of the molecular junction studied in this work. The straight lines between dots (lead atoms) depict charge electron hoppings in the lead bulks ($t'$) and between lead and molecule ($t$). The broken lines between dots (lead atoms) depict springs in the lead bulks (with elastic constant $k'$) and between lead and molecule (with elastic constant $k$). The Hot Left Lead and the Cold Right Lead are kept at chemical potential $\mu_L= eV/2$, temperature $T_L=T+\Delta T/2$ and chemical potential $\mu_R=-eV/2$, temperature $T_R=T-\Delta T/2$, respectively, with $e$ modulus of the electron charge, $V$ bias potential,  $\mu=0$ average chemical potential, $T$ average temperature. The term $U$ indicates the presence of electron-electron interactions, while the term $E_P$ electron-vibration interactions on the molecule.}
\label{fig0}
\end{figure}

The Hamiltonian $\hat{H}$ is given by
\begin{equation}
\hat{H}={\hat H}_{el}+ {\hat H}_{ph} + {\hat H}_{int},
\label{Htot}
\end{equation}
where the Hamiltonian ${\hat H}_{el}$ takes into account the electronic degrees of freedom of the leads and the molecule, 
${\hat H}_{ph}$ the vibrational degrees of freedom of the leads and the molecule, and ${\hat H}_{int}$ the coupling between electronic and vibrational degrees of freedom (see Fig. \ref{fig0} for a sketch of the molecular junction model). 

The electronic Hamiltonian ${\hat H}_{el}$ of Eq. (\ref{Htot}) is
\begin{eqnarray}
\hat{H}_{el} &=&\epsilon \sum_{\sigma} {\hat n}_{\sigma}  + U {\hat n}_{\uparrow} {\hat n}_{\downarrow}+
\sum_{q,\alpha,\sigma} \varepsilon_{q,\alpha} {\hat n_{q,\alpha,\sigma}}+  \nonumber \\
&& \sum_{q,\alpha,\sigma} \left( V_{q,\alpha}{\hat c^{\dag}_{q,\alpha,\sigma}}{\hat d}_{\sigma}+ h.c. \right),
\end{eqnarray}
where the molecular electronic level has energy $\epsilon$, the $\sigma$ spin electron density operator is ${\hat n}_{\sigma}={\hat d^{\dag}}_{\sigma}{\hat d}_{\sigma}$, with  ${\hat d^{\dag}}_{\sigma} ({\hat d}_{\sigma})$ creation (annihilation) $\sigma$ spin electron operator on the molecule. The presence of a gate in the junction can be simply simulated by changing the value of the local energy $\epsilon$. \cite{cuevas}  The Coulomb repulsion on the molecule is simulated with a Hubbard term $U$, which gives an energy penalty for electron occupations with spin $\uparrow$ and $\downarrow$. \cite{cuevas} The lead density operator is ${\hat n_{q,\alpha,\sigma}}={\hat c^{\dag}_{q,\alpha,\sigma}} {\hat c}_{q,\alpha,\sigma}$, where the operators ${\hat c^{\dag}_{q,\alpha,\sigma}} ({\hat c}_{q,\alpha,\sigma})$ create (annihilate) electrons with momentum $q$, spin $\sigma$, and energy
$\varepsilon_{q,\alpha}=\xi_{q,\alpha}-\mu_{\alpha}$ in the left ($\alpha=L$) or right ($\alpha=R$) free metallic leads, with $\mu_{\alpha}$ chemical potential of the lead $\alpha$ in equilibrium at the temperature $T_{\alpha}$. We consider the temperatures $T_L=T+\Delta T/2$ and $T_R=T-\Delta T/2$, with $T$ average temperature. Moreover, we fix the chemical potentials $\mu_L=e V/2$ and $\mu_R=-e V/2$,  with $e$ modulus of the electron charge, $V$ bias potential, and average chemical potential $\mu=0$. The electronic tunneling between the molecular dot and a state $q$ in the lead $\alpha$ has the amplitude $V_{q,\alpha}$.  As usual for metallic leads, the density of states $\rho_{q,\alpha}$ is assumed flat around the small energy range relevant for the molecular orbital, making valid the wide-band limit: $ \rho_{q,\alpha} \mapsto \rho_{\alpha}$, $V_{q,\alpha} \mapsto V_{\alpha} $. Therefore, the full hybridization width of the molecular orbital is
$\hbar \Gamma=\sum_{\alpha } \hbar \Gamma_{\alpha}$, with $\hbar$ Planck constant and the tunneling rate $\Gamma_{\alpha}=2\pi\rho_{\alpha}|V_{\alpha}|^{2}/\hbar$. In the following, we consider the symmetric configuration: $\Gamma_L=\Gamma_R=\Gamma/2$.  In junctions with $C_{60}$ molecules, $\hbar \Gamma$  has been estimated to be of the order of $20$ meV. \cite{natelson,mra} Even if the local Coulomb repulsion is reduced by the screening of the electrodes, the energy $U$ is expected to be at least one order of magnitude larger than $\hbar \Gamma$. \cite{natelson,mra}

The center of mass mode can be considered as the relevant vibrational mode of the molecule. \cite{perroni2} Indeed, experiments  have evidenced a coupling between the center of mass mode and the electron dynamics in junctions with $C_{60}$ molecules. \cite{Park} In Eq.(\ref{Htot}), the Hamiltonian ${\hat H}_{ph}$ describes the vibrations of the slow center of mass mode, the free phonon modes of the leads, and the coupling between them:
\begin{equation}
{\hat H}_{ph}={\hat H}_{cm} + \sum_{q,\alpha} \hbar \omega_{q,\alpha}{\hat a^{\dag}_{q,\alpha}}{\hat a_{q,\alpha}}
+ \sum_{q,\alpha} \left( C_{q,\alpha}{\hat a_{q,\alpha}}+h.c. \right) {\hat x}.\label{Hphon}
\end{equation}
The center of mass hamiltonian ${\hat H}_{cm}$ is
\begin{equation}
{\hat H}_{cm}={{\hat p}^{2}\over 2M} + \frac{k {\hat x}^{2}}{2},\label{Hcm}
\end{equation}
where ${\hat p}$ and ${\hat x}$ are the center of mass momentum and position operators, respectively, $M$ is the total large mass, $k$ is the effective spring constant, with frequency $\omega_{0}=\sqrt{k/M}$.
In Eq.(\ref{Hphon}), the operators ${\hat a^{\dag}_{q,\alpha}} ({\hat a}_{q,\alpha})$
create (annihilate) phonons with momentum $q$ and frequency
$\omega_{q,\alpha}$ in the lead $\alpha$. The left and right phonon leads will be considered as thermostats in equilibrium at the same temperatures $T_L$ and $T_R$, respectively, of the electron leads.
Finally, in Eq.(\ref{Hphon}), the coupling between the center of mass position and a phonon $q$ in the lead $\alpha$ is given by the elastic constant $C_{q,\alpha}$.  For large molecules, the center of mass mode has a low frequency
$\omega_{0}$ which is typically smaller than the Debye frequency $\omega_D$ of the metallic leads ($ \hbar \omega_D \simeq 15-20$ meV for metals like silver, gold, and platinum\cite{kittel}). For example, $\hbar \omega_0 \simeq 5$ meV in $C_{60}$ junctions \cite{Park}, hence $\omega_0 \simeq 0.25 \Gamma$. 
Therefore, for large molecules,  the adiabatic regime is valid for the center of mass oscillator: $\omega_0  << \Gamma$ and $\omega_0 << \omega_D$. Within this  regime, the effect of the $\alpha$ phonon lead on the center of mass mode provides a constant damping rate $\gamma_{\alpha}$.  \cite{weiss} In analogy with the electronic model, we consider the symmetric configuration: $\gamma_L=\gamma_R=\gamma/2$. For junctions with $C_{60}$ molecules and leads of Ag, Au, and Pt, $\hbar \gamma \simeq 3-8$ meV, therefore $\gamma$ is of the same order of $\omega_0$ ($\gamma \simeq 0.15-0.40 \Gamma$) \cite{perroni2}.  

Finally, the interaction term ${\hat H}_{int}$ in the Anderson-Holstein model of Eq.(\ref{Htot}) is provided by a linear coupling between the total electron density on the molecule, ${\hat n}=\sum_{\sigma} {\hat n}_{\sigma} $, and the $\hat{x}$ operator of the center of mass:
\begin{equation}
{\hat H}_{int}=\lambda {\hat x} {\hat n},\label{Hint}
\end{equation}
where $\lambda$ is the electron-vibration coupling constant. In the following, the electron-vibration interaction will be described in terms of the coupling energy $E_{P}=\lambda^2/(2 k)$.

In this paper, $\hbar \Gamma \simeq 20$ meV will be the energy unit ($\Gamma$ the frequency unit,  $1/\Gamma$ the time unit).  We will measure lengths in units of $2 \lambda / k$, and temperatures in units of $\hbar \Gamma /k_B$, with $k_B$ Boltzmann constant (the room temperature is of the order of $1.25$ in these units).

\section{Adiabatic approach within the Coulomb blockade regime}
The focus of this paper is on charge and heat transport properties close to room temperature, therefore for parameters appropriate to the Coulomb blockade regime: $\hbar \omega_0 \ll \hbar \omega_D \simeq \hbar \Gamma \leq k_B T \ll U$,  with $U > 10 \hbar \Gamma $. Besides, the electron-vibration coupling is not weak, but it is estimated to be in the intermediate regime: 
$\hbar \omega_0 \le E_P \simeq \hbar \Gamma $. Since $\hbar \omega_0$ is the lowest energy scale, the dynamics of the slow center of mass can be treated as classical. In the following, the position and the momentum of the oscillator will be indicated by the c-numbers $x$ and $p$, respectively. The parameter regime appropriate to these junctions requires a generalization of the adiabatic approach to the physical situation where the Coulomb interaction is finite and large. Recently, the adiabatic approach has been combined with a treatment of electron-electron interactions within  a slave-boson approach \cite{schiro} which is valid only in the limit of infinite local Coulomb repulsion for energies close to the chemical potential and low temperatures \cite{perroni3}.

\subsection{Electron dynamics dependent on oscillator parameters}

The electronic dynamics turns out to be equivalent to that of an adiabatically slow level with energy $E_{0}(t)=\epsilon+\lambda x(t)$ within the Coulomb blockade regime. \cite{fazio,mucciolo} 

At the zero order of the adiabatic expansion, the electronic quantities can be calculated considering an energy level with a fixed oscillator position $x$.  The effects of the strong Coulomb repulsion are treated inserting the first self-energy correction upon the atomic limit. \cite{Haug}  Therefore, for the paramagnetic solution, the level spectral function $A_0(\omega,x)$ at zero order of the adiabatic expansion becomes
\begin{eqnarray}
A_0(\omega,x) &=& [1-\rho(x)]\frac{\hbar \Gamma}{(\hbar \omega - \epsilon -\lambda x )^2+(\hbar \Gamma)^2/4}+  \nonumber \\
&& \rho(x) \frac{\hbar \Gamma}{(\hbar \omega - \epsilon -\lambda x -U )^2+(\hbar \Gamma)^2/4},
\label{function}
\end{eqnarray}
where $\rho(x)$ is the level density per spin self-consistently calculated at fixed position $x$ through the following integral 
\begin{equation}
\rho(x)=\int_{-\infty}^{+\infty} \frac{d (\hbar \omega)}{2 \pi i} G^<_0(\omega,x),
\end{equation}
with the lesser Green function $G^<_0(\omega,x)$
\begin{equation}
G^<_0(\omega,x)=\frac{i}{2} [f_L(\omega)+f_R(\omega)] A_0(\omega,x),
\end{equation}
and $f_{\alpha}(\hbar \omega)=1/(\exp{[\beta_{\alpha} (\hbar \omega-\mu_{\alpha})]}+1)$ Fermi distribution of the lead $\alpha$ ($\beta_{\alpha}=1/k_B T_{\alpha}$). Actually, the spectral function is characterized by a double peak structure that, for large $U$, is robust against the effects of electron-vibration coupling which tend to shift and enlarge the single peaks (the single peak width increases by a factor of the order of $E_P$).  

In Appendix A, we compare the spectral function of this treatment for strong Coulomb repulsion with that of another approach which retains additional self-energy corrections upon the atomic limit in the absence of electron-vibration coupling. \cite{Haug} For large $U$ and room temperature, the approach considered here is very accurate, therefore, it represents an optimal starting point for the adiabatic expansion. 
In this paper, we will study different properties varying the electronic level occupation. In our model, these variations can be controlled changing the molecule level energy $\epsilon$ with respect to the leads chemical potential (average chemical potential $\mu=0$ in this work). In Appendix A, we report the molecular electron occupation $N$ as a function of  level energy $\epsilon$ showing the typical profiles of the Coulomb blockade. In particular, the following energies are relevant: $\epsilon=-U/2$ (close to half-filling $N=1$), $\epsilon=-U$ (transition from level occupation $N=1$ to $N=2$), $\epsilon=0$ (from level occupation $N=1$ to $N=0$). 

Within the adiabatic approach, one can determine the electronic Green functions and generic electronic quantities making an expansion on the small oscillator velocity $v=p/m$. In the absence of electron-electron interactions, the adiabatic expansion can be determined for any strength of electron-vibration coupling. \cite{alberto,alberto1,alberto2,perroni,perroni1} In this paper, an approach is devised for the case of strong Coulomb repulsion in order to include the effects of electron-vibration interaction within the realistic intermediate coupling regime.  Actually, the approach used in this paper is valid as long as the two peaks characteristic of Coulomb blockade can be resolved, therefore for the physical regime $E_P \ll U$. In the next subsection, we will use the adiabatic expansion of the level occupation to derive the motion equation of the slow center of mass oscillator in a self-consistent way.

\subsection{Dynamics of the center of mass oscillator}

The effect of the molecule electron degrees of freedom and of the phonon baths in the leads gives rise to the following generalized Langevin equation for the slow center of mass 
\begin{equation}
m \frac{d v}{d t}= F_{det}(x,v) + \xi(x,t),
\label{langevin1}
\end{equation}
which has the deterministic force $F_{det}(x,v)$ and the position dependent fluctuating force $\xi(x,t)$. The deterministic force
\begin{equation}
F_{det}(x,v)=F_{gen}(x)-A_{eff}(x) v,
\label{fortot1}
\end{equation}
can be decomposed into a generalized force $F_{gen}(x)$
\begin{equation}
F_{gen}(x)=-k x +F_{\lambda}(x),
\label{fgen}
\end{equation}
with $F_{\lambda}(x)=-2 \lambda  \rho(x)$ induced by the electron-vibration coupling, and, as a result of the adiabatic expansion, 
a dissipative force with an effective position dependent positive definite term $A_{eff}(x)$
\begin{equation}
A_{eff}(x)=A_{\lambda}(x)+m \gamma,
\label{Aeff}
\end{equation}
with $A_{\lambda}(x)$ 
\begin{equation}
A_{\lambda}(x)=2 \hbar \lambda^2 \int_{-\infty}^{+\infty} \frac{d (\hbar \omega)}{2 \pi i} G^<_0(\omega,x) 
\left[ \frac{\partial A_0(\omega,x)}{\partial (\hbar \omega)} \right]
\label{Alam}
\end{equation}
due to the electron-vibration interaction. The fluctuating force $\xi(x,t)$ in Eq.(\ref{langevin1}) is such that
\begin{equation}
\langle \xi(x,t) \rangle=0,\;\;\;\; \langle \xi(x,t) \xi(x,t') \rangle= D_{eff}(x) \delta(t-t'), \nonumber
\label{Langevin20}
\end{equation}
where the effective position dependent noise term $D_{eff}(x)$ is
\begin{equation}
D_{eff}(x)=D_{\lambda}(x)+ k_B (T_L+T_R) m \gamma,
\end{equation}
with $D_{\lambda}(x)$ 
\begin{equation}
D_{\lambda}(x)=2 \hbar \lambda^2 \int_{-\infty}^{+\infty} \frac{d (\hbar \omega)}{2 \pi i} G^<_0(\omega,x)
G^>_0(\omega,x)
\label{Dlam}
\end{equation}
determined by the electron-vibration coupling and the greater Green function $G^>_0(\omega,x)$
\begin{equation}
G^>_0(\omega,x)=-\frac{i}{2} [2-f_L(\omega)-f_R(\omega)] A_0(\omega,x).
\end{equation}
It is worthwhile pointing out that, in equilibrium conditions at temperature $T=T_{\alpha}$ and chemical potential $\mu=\mu_{\alpha}=0$, the adiabatic procedure gives rise to a generalized fluctuation-dissipation relation $D_{eff}(x)=2 k_B T A_{eff}(x)$ valid for each fixed position $x$.  

The solution of the Langevin equation (\ref{langevin1}) represents a central step for this work. 
This equation has been numerically solved under generic non-equilibrium conditions using a generalized Runge-Kutta algorithm. \cite{alberto,honey,honey1} As a result of the numerical calculations, the oscillator distribution function $Q(x,v)$ and the reduced position distribution function $P(x)$ are determined allowing to evaluate static quantities relative to the center of mass oscillator. Moreover, these distribution functions will allow to make the average of an electronic observable $O(x,v)$ dependent on oscillator parameters.

\begin{figure}
\centering
\includegraphics[width=8.5cm,height=8.5cm]{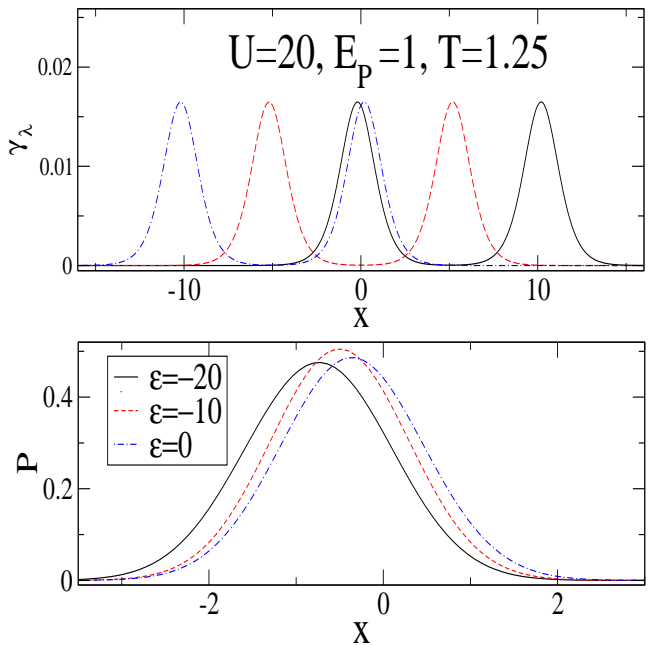}
\caption{(Color online) Electron-vibration induced damping rate $\gamma_{\lambda}$  in units of $\Gamma$ (Upper Panel) and reduced position distribution function $P$ in units of $k/ 2 \lambda $ (Lower Panel) as a function of oscillator position x (in units of $ 2 \lambda / k$) for different values of level energy $\epsilon$ (in units of $\hbar \Gamma$). In all the plots, $U=20 \hbar \Gamma$, $E_P=  \hbar \Gamma$, and temperature $T=1.25 \hbar \Gamma /k_B $ (close to room temperature).}
\label{fig1}
\end{figure}

Before going to the section about results, we discuss the features of the electron-vibration induced damping rate $\gamma_{\lambda}(x) =A_{\lambda}(x)/m$, with $A_{\lambda}(x)$ given in Eq.(\ref{Alam}).  The magnitude of $\gamma_{\lambda}(x)$ always gets enhanced with increasing the electron-vibration coupling $E_P$. However, as reported in the upper panel of Fig. \ref{fig1}, even for the intermediate coupling $E_P=1$, the peak values of $\gamma_{\lambda}(x)$ are always smaller than realistic values of the lead induced damping rate $\gamma$ ($\gamma = 0.15$ will be considered in this paper). This implies that the effects due to the electron-vibration coupling on the oscillator dynamics do not typically represent a large perturbation with respect to those induced by the coupling to phonon leads. Obviously, as reported in the figure, the behavior of $\gamma_{\lambda}(x)$  strongly depends on the occupation of the electronic level. We point out that, in contrast to the spinless case analyzed in a recent paper, \cite{perroni2} $\gamma_{\lambda}(x)$ shows a double-peak behavior due to the effect of the strong Hubbard interaction. Moreover, as reported in the upper panel of Fig. \ref{fig1}, the peaks of 
$\gamma_{\lambda}(x)$ largely shift passing from the quasi half-filling case (close to $\epsilon=-10=-U/2$,  state with flat occupation) to conditions out of half-filling (close to $\epsilon=-20=-U$ and $\epsilon=0=\mu$, state with strong density fluctuations).  The self-consistent calculation of $\gamma_{\lambda}(x)$ provides a direct signature of the strong local interaction since it is determined by the adiabatic expansion of the electron occupation.

A comparison of the $x$ dependence between $\gamma_{\lambda}(x)$ and the calculated oscillator position distribution $P(x)$ will clarify the conditions under which the electron-vibration interaction can affect the dynamics of the center of mass oscillator. Therefore, in the lower panel of Fig. \ref{fig1}, we report the  distribution $P(x)$ with varying the level energy $\epsilon$. We notice that, apart from the shift of the peaks, close to room temperature, the distribution $P(x)$ is practically the Gaussian of the free harmonic oscillator at temperature T for any value of the level energy $\epsilon$. In the quasi half-filled case ($\epsilon=-10$), the peak positions of $\gamma_{\lambda}(x)$ and $P(x)$ are well separated. Therefore, one expects that, in this regime, the effects of the electron-vibration coupling on the oscillator dynamics are weak. We stress that, within the self-consistent procedure used in this work, the peak of the $P(x)$ directly signs the level occupation being close to $-N/2$ within the units used in this paper. Actually,  for $\epsilon=-10$, the value close to $-0.5$ of the peak of  $P(x)$ is fully compatible with the half-filled case $N=1$. On the other hand, for $\epsilon=-20$, the peak position of $P(x)$ shifts towards lower values close to  $-0.75$ ($N \simeq 1.5$), and, for $\epsilon=0$, to $0.25$ ($N \simeq 0.5$). We point out that, for 
$\epsilon=-20$, the first peak of $\gamma_{\lambda}(x)$ is close to $x=0$, while, for $\epsilon=0$, the second peak of 
$\gamma_{\lambda}(x)$  strongly overlaps with the position distribution $P(x)$. Therefore, out of half-filling, the effects of the electron-vibration coupling can affect the oscillator dynamics. In contrast with the spinless case, \cite{perroni2}  these effects are present not only close to $\epsilon=\mu=0$, but also to $\epsilon=-U=-20$, as a result of the strong Coulomb interaction. Therefore, as discussed in detail in the next section, the complex interplay between electron-electron and electron-vibration interactions opens an entire energy region where the phonon heat transport can be enhanced.

\section{Results}
In this paper, we will discuss linear response transport properties  trying to clarify the role of 
the electron-electron and electron-vibration interactions. In the next subsections, we will analyze the phonon heat transport, the electronic spectral function, the charge and electronic heat transport, and thermoelectric figure of merit.  In the following, we will assume $\omega_0=0.25 \Gamma$, and $\gamma = 0.15 \Gamma$ (larger values of  $\gamma$ have been considered in a recent paper \cite{perroni2}).

\subsection{Phonon heat transport}

In this subsection, we will focus on the phonon thermal conductance $G_K^{ph}$ calculated within the linear response regime around temperature T  as
\begin{equation}
G_K^{ph}=\lim_{\Delta T \rightarrow 0^+} \frac{(J_L^{ph}-J_R^{ph})}{2 \Delta T},
\end{equation}
with $J_{\alpha}^{ph}$ current from the $\alpha$ phonon lead. \cite{wang3,perroni2} 

\begin{figure}
\centering
\includegraphics[width=9.5cm,height=7cm]{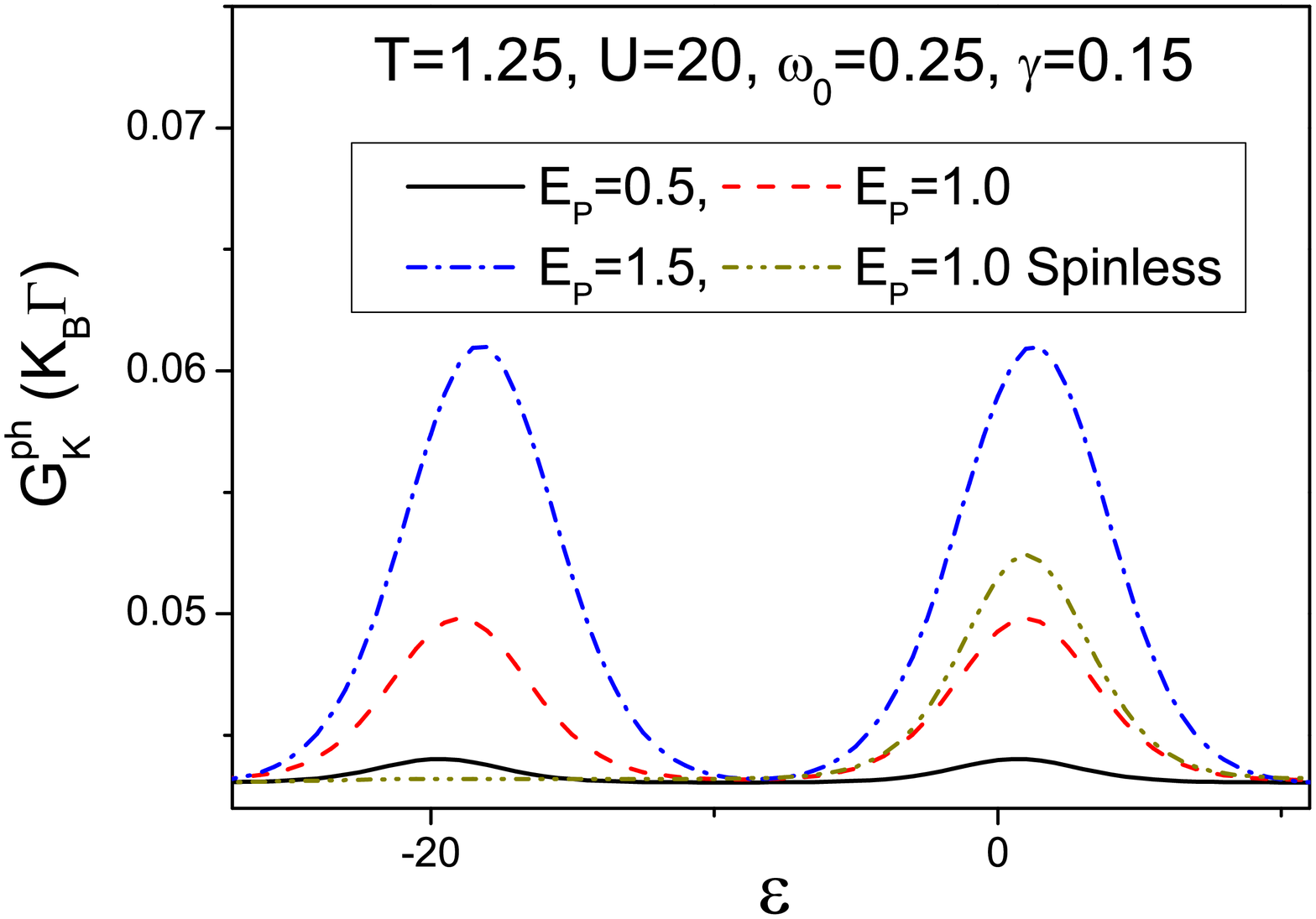}
\caption{(Color online) Phonon thermal conductance $G_K^{ph}$ (in units of $k_B \Gamma$) as a function of the level energy $\epsilon$ (in units of $\hbar \Gamma$) for different values of electron-vibration coupling $E_P$ (in units of $\hbar \Gamma$).
In the plot, $U = 20 \hbar \Gamma $, $T=1.25 \hbar \Gamma \ k_B$ (close to room temperature), $\omega_0=0.25 \Gamma$, and oscillator damping rate $\gamma=0.15 \Gamma$.}
\label{fig2}
\end{figure}

The conductance $G_K^{ph}$ is expected to be mostly sensitive to the coupling of the center of mass mode to the phonons of metallic leads  through the damping rate $\gamma$ ($\gamma  = 0.15 \Gamma$ in this work) which is typically larger than the peak values of 
electron-vibration induced damping rate $\gamma_{\lambda}(x)$.  As shown in Fig.\ref{fig2}, in the regime of weak electron-vibration coupling $E_P$, low level occupation ($\epsilon \gg 0$), and double level occupation ($\epsilon \ll -U$), $G_K^{ph}$ is close to $0.04$ $k_B \Gamma$ ($k_B \Gamma$ is about $419.8$ pW/K for $\hbar \Gamma \simeq 20$ meV), a numerical value coincident with an analytical estimate of $G_K^{ph}$ given in a recent paper \cite{perroni2}. This asymptotic value corresponds to the contribution given by the only phonon leads neglecting the effects of electron-electron and electron-vibration interactions on the molecule. 

In Fig. \ref{fig2}, we show that $G_K^{ph}$ always gets larger with increasing the electron-vibration coupling $E_P$. Moreover, this increase of  $G_K^{ph}$ strongly depends on the value of level energy $\epsilon$. In contrast with the spinless case (reported for comparison in Fig. \ref{fig2} at $E_P=1$), we stress that the enhancement of $G_K^{ph}$ takes place not only close to $\epsilon \simeq 0$, but also to $\epsilon \simeq -U$.  Therefore, the distance between the peaks of the phonon thermal conductance is controlled by the energy scale $U$. The peak values are almost coincident (although slightly smaller than the peak value of the spinless case), and, at $E_P=1$, they are of the order of $0.05 k_B \Gamma \simeq 20$ pW/K. Therefore, the calculated $G_K^{ph}$ is in very good agreement  with the thermal conductance of the order of a few $10$ pW/K measured for molecules anchored to gold \cite{wang2,meier}. In any case, due to the strong electron-electron interactions, $G_K^{ph}$ can be enhanced in a new large energy region. On the other hand, for 
$\epsilon \simeq -U/2$, $G_K^{ph}$  is poorly influenced by the electron-vibration effects even if $E_P$ is not small getting a value close to the asymptotic one.  From this analysis emerges that the complex enhancement of the phonon thermal conductance $G_K^{ph}$ as a function of the electron-electron and electron-vibration interactions can be mostly ascribed to the properties of additional electron-vibration induced damping rate $\gamma_{\lambda}(x)$ discussed in the previous section.  

\subsection{Electronic spectral function}

From the solution of the Langevin equation, one can make the average of an electronic observable $O(x,v)$ over the oscillator distribution function. First, we discuss the features of the electronic spectral function which is at the basis of the thermoelectric properties analyzed in the next subsection.

The electronic spectral function $A(\omega)$ is evaluated making the average of the function $A_0(\omega,x)$ in Eq.(\ref{function}) over $P(x)$:
\begin{equation}
A(\omega) =\int_{-\infty}^{+\infty} d x P(x) A_0(\omega,x).
\label{spec}
\end{equation}
In this section, the spectral function will be discussed in equilibrium conditions at temperature $T$ ($V=0$ and $\Delta T=0$). We recall that, in Appendix A, the features of the spectral function are discussed in the absence of electron-vibration coupling.  Actually, the spectral function is characterized by a structure with two peaks separated by an energy of the order of $U$, and it is strongly dependent on the value of the level energy $\epsilon$. 

\begin{figure}
\centering
\includegraphics[width=8cm,height=9cm]{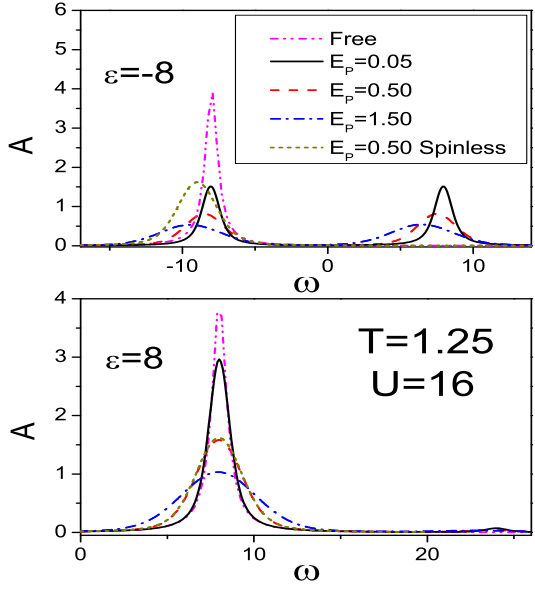}
\caption{(Color online) Spectral function (in units of $1/ \hbar \Gamma$) as a function of frequency $\omega$ 
(in units of $\Gamma$) at level energy $\epsilon=-8 \hbar \Gamma$ (Upper Panel) and $\epsilon=8 \hbar \Gamma$ (Lower Panel) for different values of $E_P$ (in units of $\hbar \Gamma$).
In all the plots, $U=16 \hbar \Gamma$, $T=1.25 \hbar \Gamma / k_B$ (close to room temperature).}
\label{fig3}
\end{figure}

In this subsection, we analyze the behavior of the spectral function with varying the electron-vibration coupling $E_P$ at a fixed value of Hubbard energy $U$. In the upper panel of Fig. \ref{fig3},  we show the spectral function for different values of the electron-vibration coupling in the half-filled case $\epsilon=-8=-U/2$ (level occupation $N=1$). For comparison, we report the spectral function relative to the case where electron-electron and electron-vibration interactions are neglected (indicated as Free in the figure). 
We point out that there is a strong transfer of spectral weight for the double peak structure toward low frequencies with increasing $E_P$. In addition to the shifts of the peaks, the electron-vibration coupling tends to reduce the height of the peaks and to enlarge them. Actually, the single peaks increase their width by a factor of order of $E_P$. We stress that, for realistic values of the coupling $E_P$, the two Hubbard peaks do not overlap, therefore the double peak structure due to the large $U$ is quite robust to the effects of electron-vibration coupling. Finally, we notice that, in the spinless case (reported for comparison in Fig. \ref{fig3} at $E_P=0.5$), the spectral function has a single peak, and it is quite sensitive to the effects of the electron-vibration coupling.   

As shown in the lower panel of Fig. \ref{fig3}, a different behavior takes place in the regime of low level occupation ($\epsilon=8$ in the figure). For the considered values of $E_P$, the spectral function gets enlarged, but its peak position is quite rigid. 
Moreover, the differences with the spinless case are completely negligible. Even in the presence of electron-vibration coupling  $E_P$, the behavior of the spectral function is different in the regime of half-filling and of low or high level occupation.

\subsection{Charge and electronic heat transport, and thermoelectric figure of merit}

In this subsection, the focus will be on the regime of linear response around the average chemical potential $\mu=0$ and temperature T ($\Delta T \rightarrow 0^+ $, $V \rightarrow 0^+$). We will evaluate the electronic conductance $G$
\begin{equation}
G=\left( \frac{2 e^2}{\hbar} \right) \left( \frac{\hbar \Gamma}{4} \right) \int_{-\infty}^{+\infty} \frac{ d (\hbar \omega)}{2 \pi} A(\omega) \left[ -\frac{\partial f(\hbar \omega)}{\partial (\hbar \omega)} \right],
\label{conduct}
\end{equation}
where $f(\hbar \omega)=1/(\exp{[\beta (\hbar \omega-\mu)]}+1)$ is the free Fermi distribution corresponding to the average chemical potential 
$\mu=0$. Then, we will calculate the Seebeck coefficient $S=-G_S/G$, with
\begin{equation}
G_S=  \left( \frac{2 e}{\hbar} \right) \left( \frac{\hbar \Gamma}{4 T} \right) \int_{-\infty}^{+\infty}
\frac{ d (\hbar \omega)}{2 \pi} (\hbar \omega) A(\omega) \left[ -\frac{\partial f(\hbar \omega)}{\partial (\hbar \omega)} \right].
\label{conducts}
\end{equation}
Finally, we will determine the electron thermal conductance $G_K^{el}=G_Q+T G_S S$, with
\begin{equation}
G_Q=  \left( \frac{2}{\hbar T} \right) \left( \frac{\hbar \Gamma}{4 } \right) \int_{-\infty}^{+\infty}
\frac{ d (\hbar \omega)}{2 \pi} (\hbar \omega)^2 A(\omega) \left[ -\frac{\partial f(\hbar \omega)}{\partial (\hbar \omega)} \right].
\label{conductq}
\end{equation}
The total thermal conductance $G_K=G_K^{el}+G_K^{ph}$ makes feasible the evaluation of the figure of merit $ZT=G S^2 T /G_K$.
When the coupling of the center of mass mode to the metallic leads is absent ($\gamma=0$), $G_K=G_K^{el}$, so that $ZT =ZT^{el}$, which can be used to characterize the electronic thermoelectric efficiency.

\begin{figure}
\centering
\includegraphics[width=9.5cm,height=9.5cm]{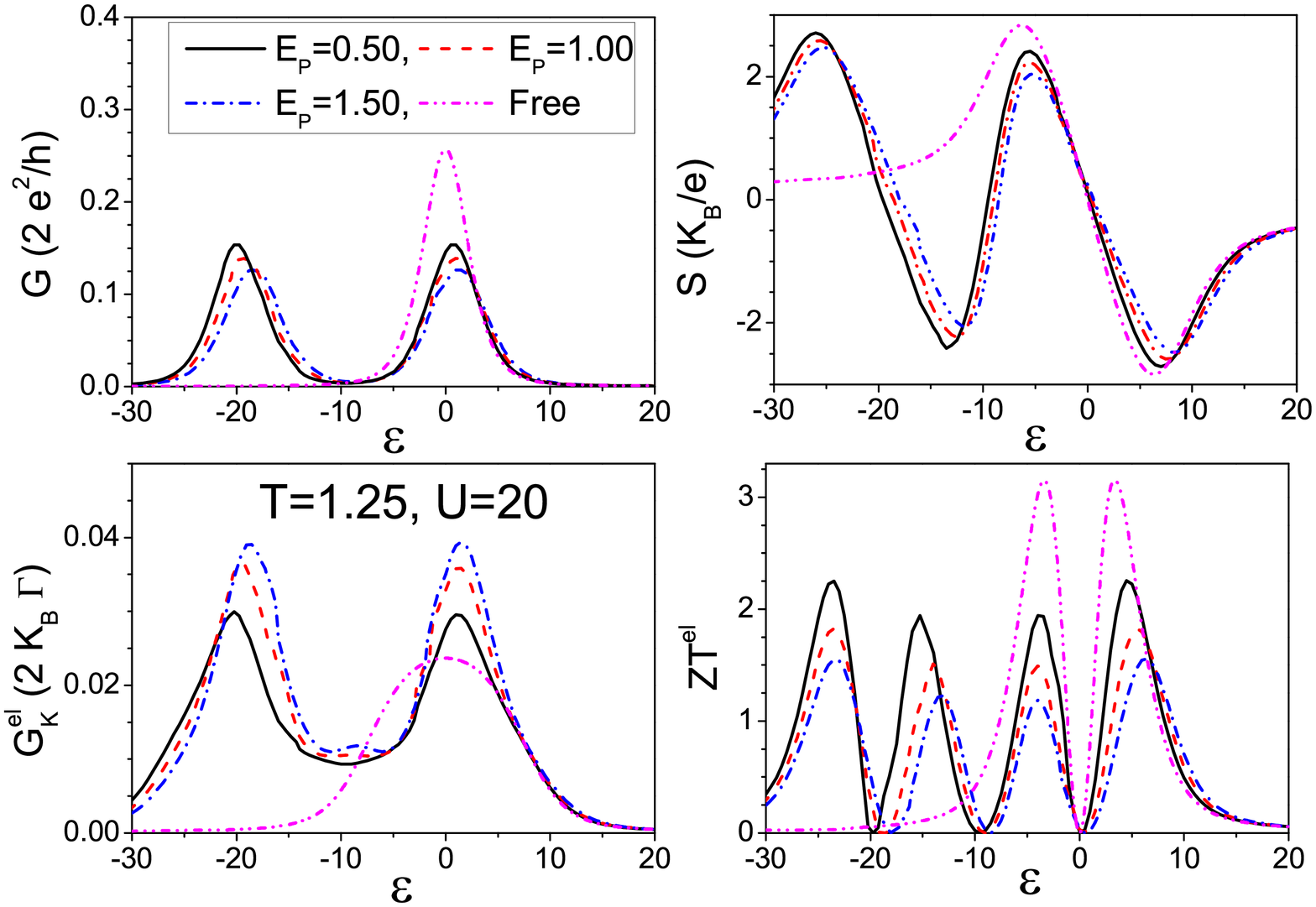}
\caption{ (Color online) Electron conductance G in units of $2 e^2/h$ (Upper Left Panel), Seebeck coefficient $S$ in units of $k_B/e$ (Upper Right Panel), electron thermal conductance $G_K^{el}$ in units of $2 k_B \Gamma$ (Lower Left Panel), and figure of merit $ZT^{el}$ (Lower Right Panel) as a function of the level energy $\epsilon$ (in units of $\hbar \Gamma$) for different values of electron-vibration $E_P$ (in units of $\hbar \Gamma $).
In all the plots, $U=20 \hbar \Gamma$, $\gamma=0$, and $T=1.25 \hbar \Gamma / k_B$ (close to room temperature). }
\label{fig4}
\end{figure}

As reported in  Fig. \ref{fig4}, we analyze the effects of the electron-vibration coupling on the electronic response functions as a function of the level energy $\epsilon$ at a fixed value of Hubbard interaction $U$ ($U=20$) in the absence of coupling to phonon leads ($\gamma=0$) close to room temperature ($T=1.25$). For comparison, we report the transport properties relative to the case when electron-electron and electron-vibration interactions are neglected (indicated as Free in the figure). 
  
The charge conductance $G$ is expected to be smaller than the free one due to the effects of interactions. As shown in the upper left panel of Fig. \ref{fig4}, close to room temperature, $G$ has peak values of the order of  $10^{-1}$ $e^2/h$ ($e^2/h$ is about $3.87 \times 10^{-5}$ S). In particular, for $\epsilon \simeq 20$, we have checked that $G$ is of the order of $10^{-3}$ $e^2/h$ in agreement with the order of magnitude of experimental data in $C_{60}$ \cite{majum2}. As expected, the conductance as a function of the level energy $\epsilon$ follows a behavior similar to the double-peak structure of the spectral function as a function of the frequency. Therefore, $G$ has maxima for $\epsilon \simeq 0=\mu$ and $\epsilon \simeq -U$, and a minimum at $\epsilon \simeq -U/2$.

As shown in the upper right panel of Fig. \ref{fig4}, the Seebeck coefficient $S$ shows large variations with changing $\epsilon$. 
Indeed, $S$ shows two maxima and two minima whose magnitude is very large at room temperature being of the order of $2$ $k_B/e$ ($k_B/e$ is about $86$ $\mu$eV/K). This complex behavior is due to the role played by the strong electron correlations \cite{liu}. Actually, the structure close to $\epsilon=0$ (where $S$ vanishes) is nearly translated by $-U$ (for $\epsilon \simeq -20$, $S$ goes again to zero). Therefore, even at $\epsilon \simeq -U/2$, $S$ gets very small values.  Obviously, for large positive values of $\epsilon$, $S$ is small and negative (n-type behavior).  In particular, for $\epsilon=20$, $S$ is about $-0.45 k_B/e \simeq - 38.5 \mu V/K$ in agreement with the magnitude of experimental data in $C_{60}$ \cite{majum2}.  


As shown in the upper panels of Fig. \ref{fig4}, the most relevant effect of the coupling $E_P$ on the conductance $G$ and the Seebeck coefficient $S$ is to shift the curves and reduce the magnitude of the response function. The shift of the conductance peaks and of the zeroes of the Seebeck coefficient is of the order of $E_P$. At fixed level energy, unlike the conductance $G$, the Seebeck coefficient is more sensitive to the changes of the coupling $E_P$. For example, this occurs for energies close to the minima and the maxima. By changing the values of $\epsilon$, there is an inversion in the behavior of $S$ with increasing the electron-vibration coupling $E_P$.

As shown in the lower left panel of Fig.\ref{fig4}, with varying the level energy $\epsilon$, the electron thermal conductance $G_K^{el}$ shows the characteristic double peak structure due to correlation effects \cite{liu}. The peaks values of $G_K^{el}$ are of the order of a few $0.01$ $k_B \Gamma$ ($k_B \Gamma$ is about $4.198 \times 10^{-10}$ W/K for $\hbar \Gamma \simeq 20$ meV). Therefore, the peak values are smaller than  the thermal conductance quantum $g_0(T) = \pi^2 k_B^2 T/(3 h)$ at the room temperature $T=1.25 \hbar \Gamma \simeq 300 $K ($g_0(T) \simeq  9.456 \times 10^{-13} (W/K^2) T)$ \cite{jezouin}.   
We point out that electron-vibration interactions affect  the thermal conductance $G_K^{el}$ in a way completely different from the charge conductance $G$ (compare left upper and left lower panel of  Fig. \ref{fig4}). Indeed, $G_K^{el}$ gets enhanced with increasing the electron-oscillator coupling $E_P$. As discussed in the previous section, within the adiabatic approach,  the molecular effective level is renormalized by the position variable $x$ which has a larger spreading upon increasing the electron-vibration coupling.

We stress that the behavior of the electron thermal conductance $G_K^{el}$ shown in the lower left panel of  Fig. \ref{fig4} bears a strong resemblance with that of the phonon thermal conductance $G_K^{ph}$  reported in Fig. \ref{fig2}. Both have a double peak structure, and both are enhanced by the electron-vibration coupling. Moreover, $G_K^{el}$ acquires values larger than those of $G_K^{ph}$ in the energy region $-U \le \epsilon \le 0$. Obviously, the values of these quantities are comparable for the chosen value of phonon induced damping rate $\gamma= 0.15 \Gamma$. If one consider larger values of $\gamma$ (for example $\gamma \simeq 0.4 \Gamma$), then $G_K^{ph}$ would play a major role in the total thermal conductance $G_K$.  In any case, the values of $G_K^{ph}$ and $G_K^{el}$  differ for $\epsilon \gg 0$ and $\epsilon \ll -U$ since $G_K^{ph}$ acquires a finite asymptotic value (obtained even in the absence of interactions on the molecule), while $G_K^{el}$ goes rapidly to zero.   

As shown in the lower right panel of Fig. \ref{fig4}, we analyze the behavior of the electronic thermoelectric figure of merit $ZT^{el}$  neglecting the contribution from $G_K^{ph}$. The quantity $ZT^{el}$ shows four peaks whose values are larger than $1$, but smaller than the peak value around $3$ obtained in the absence of interactions. We stress that the peak values of $ZT^{el}$ at room temperature are almost coincident with maxima and minima of the Seebeck coefficient $S$. Actually, close to room temperature, the small values of the conductance $G$ are fully compensated by the large values of the Seebeck coefficient $S$. With increasing the electron-vibration coupling $E_P$, the reduction of $G$ and $S$ combines with the enhancement of $G_K^{el}$ leading to a sensible reduction of the figure of merit $ZT^{el}$. Therefore, even if one neglects the role of phonon thermal conductance, the effect of electron-electron and electron-vibration interactions is able to induce a reduction of the figure of merit.

\begin{figure}
\centering
\includegraphics[width=9.0cm,height=7cm]{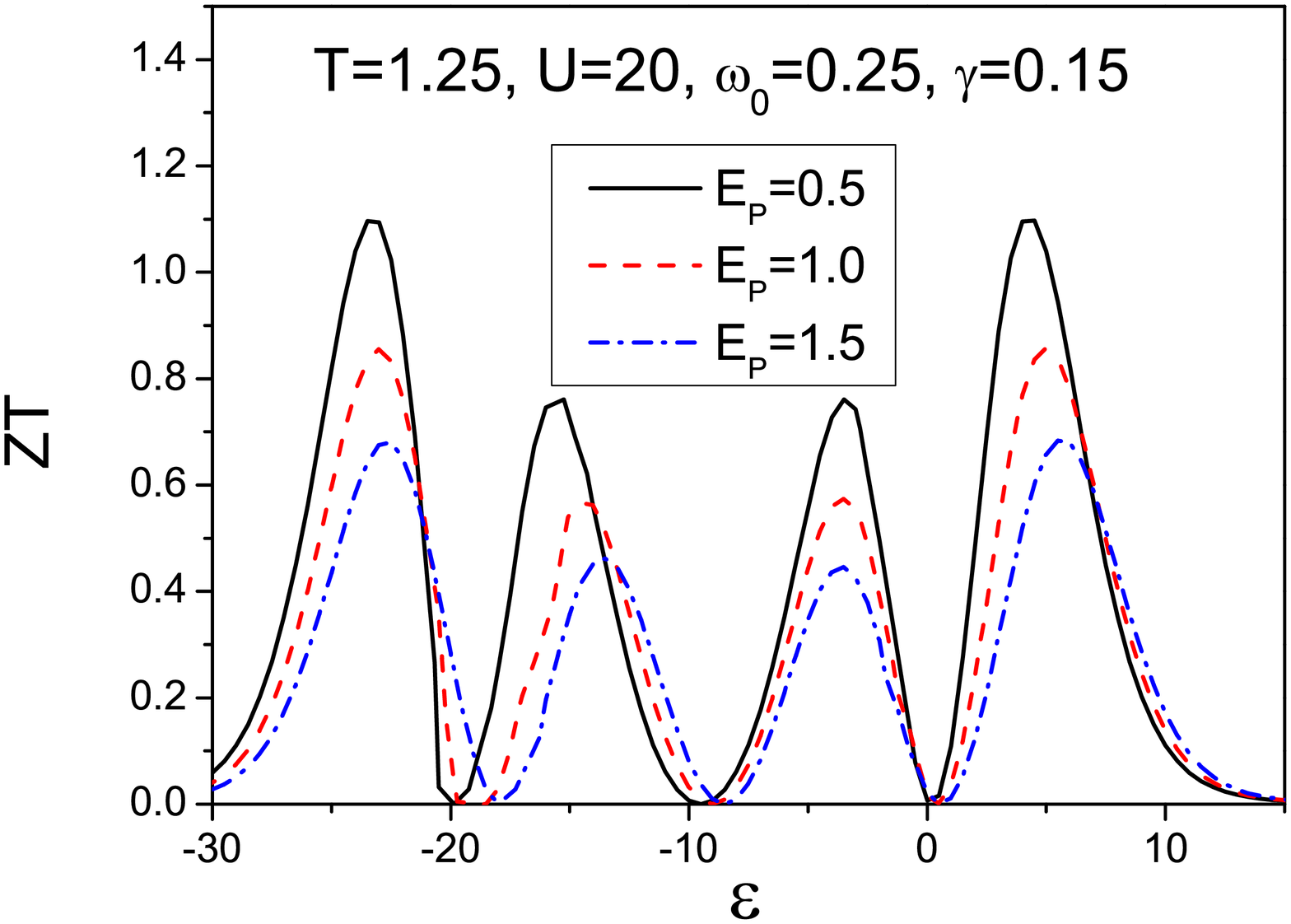}
\caption{(Color online) Dimensionless thermoelectric figure of merit $ZT$ as a function of level energy $\epsilon$ (in units of $\hbar \Gamma$) for different values of electron-vibration coupling $E_P$ (in units of $\hbar \Gamma$). In the plot, $U=20 \hbar \Gamma$, $T=1.25 \hbar \Gamma / k_B$, $\gamma=0.15 \Gamma$ and $\omega_0=0.25 \Gamma$.}
\label{fig5}
\end{figure}

In Fig. \ref{fig5}, we focus on the total figure of merit $ZT$ as a function of the level energy $\epsilon$ for different values of electron-vibration coupling $E_P$ at $U=20  \hbar \Gamma$ including the effects of the phonon thermal conductance ($\gamma=0.15 \Gamma$). From the comparison with the results discussed in the previous paragraph, it emerges that the phonon thermal conductance $G_K^{ph}$  induces an additional suppression of $ZT$. For the realistic value of  $E_P=0.5$ (intermediate coupling regime), the peak values of $ZT$ are decreased by a factor of $2$ in comparison with $ZT^{el}$, therefore the reduction of $ZT$ is not strong. Only for unrealistically large electron-vibration couplings ($E_P$ larger than $1$), $ZT$ acquires peak values less than unity. Summarizing, the cooperative effects of phonon leads, electron-electron and electron-vibration interactions on the molecule are able to weaken the thermoelectric performance of this kind of device. However, within a realistic regime of parameters, the thermoelectric figure of merit $ZT$ is still of the order of unity, making these devices a valid choice for thermoelectric applications.

\section{Conclusions}

In this paper, the thermoelectric properties of a molecular junction have been studied within the linear response regime  at room temperature. In particular, we have analyzed the role played by the phonon thermal contribution $G_K^{ph}$ on the figure of merit $ZT$ in the presence of realistic electron-electron and electron-vibration interactions. The interplay between the low frequency center of mass oscillation of the molecule and the electronic degrees of freedom has been investigated using a non-equilibrium adiabatic approach generalized  for including the large electron-electron Coulomb repulsion. Parameters appropriate to $C_{60}$ molecules have been considered. Within the intermediate electron-vibration coupling regime, the phonon thermal conductance  $G_K^{ph}$  is quite sensitive to the changes in the occupation of electron level. Moreover, apart from an important asymptotic value,  $G_K^{ph}$ resembles the electron thermal conductance $G_K^{el}$. With increasing the electron-vibration coupling, the phonon and the electron thermal conductance get larger, while the charge conductance $G$ and the thermopower $S$ get smaller.  The figure of merit $ZT$ depends appreciably on the behavior of $G_K^{ph}$ and intramolecular interactions. Indeed, for realistic parameters of the model, $ZT$ can be substantially reduced, but its peak values can be still of the order of unity indicating that the emerging field of molecular thermoelectrics can be very interesting for applications. 

The parameters of the junction are determined by the coupling between molecule and metallic leads in the electronic and vibrational channels. For instance, the strength of the intramolecular couplings depends on the choice of the leads which screen the electron-electron and electron-vibration interactions.  In order to improve the thermoelectric efficiency, molecules and metallic leads forming the junction have to ensure a weak phonon-center of mass coupling (small $\gamma$) and a small strength of the electron-center of mass interaction (small $E_P$). For realistic values of these couplings,  the values of the phonon thermal conductance $G_K$ are small compared to bulk conductances. 
Therefore, the values of $ZT$ of the order of unity can be found in molecular junctions since these systems provide a mechanism to keep the phonon thermal conduction lower than that of bulks and other low-dimensional structures. Finally, in this paper, we have shown that, for realistic values of junction parameters, the phonon thermal conductance can be even smaller than the electron counterpart in a large range of gate voltages.

The electron-vibration interaction of the Anderson-Holstein model analyzed in this paper is related to the charge density injected by the external leads onto the molecule. The renormalization of the lead-molecule hopping integral induced by the center of mass movement could represent another possible source of electron-vibration coupling \cite{koch} and it can be studied within the adiabatic approach. However, we expect that the coupling through electron level density plays a major role due to the large mass of the molecules considered in this work. Finally, we stress that the approach proposed in this paper can be generalized to the study of more realistic multi-level molecular models and to cases where the number of atomic units within the molecule can be varied.     


\begin{acknowledgments}
This work has been performed in the frame of the project GREEN (PON02 00029 2791179) granted to IMAST S.c.a.r.l. and funded by the MIUR (Ministero dell'Istruzione, dell'Universit\`a e della Ricerca.). 
\end{acknowledgments}

\begin{appendix}

\section{Comparison between different approaches within the Coulomb blockade regime}

In this Appendix, we compare the approach used in the main text for strong Coulomb repulsion with that of Lacroix \cite{Lacroix}  which retains additional self-energy corrections upon the atomic limit \cite{Haug}.  We will consider the electronic properties in the absence of electron-vibration coupling since we are interested only on the effects induced by the electron-electron interaction in equilibrium conditions at temperature $T=T_{\alpha}$ and chemical potential $\mu=0=\mu_{\alpha}$.  In this Appendix, we will use the same units of the main text. 

In contrast with the main text, in this Appendix, we will use a slightly different kind of wide-band approximation for the electron leads. Actually, we will consider an energy dependent tunneling rate $\Gamma_0(E)=\Gamma$, for $-E_c \le E \le E_C$, and zero elsewhere, with $E_C$ cutoff energy much larger than $U$. Therefore, the retarded self-energy of the electron level $\Sigma_0(E)$ due to the effects of the electron leads is
\begin{equation}
\Sigma_0(E)=\Lambda_0(E)-\frac{i}{2} \Gamma_0(E),
\end{equation}  
where $\Lambda_0(E)$ is the real part of the retarded self-energy
\begin{equation}
\Lambda_0(E)=\int_{-\infty}^{+\infty} \frac{d E'}{2 \pi} \frac{\Gamma_0(E')}{E -E' +\mu}=\frac{\Gamma}{2 \pi} \ln{\left| \frac{E-E_C+\mu}{E+E_C+\mu} \right|}.
\end{equation} 
In the limit where $E_C \rightarrow \infty$, one recovers the wide band approximation used in the main text corresponding to a zero real part $\Lambda_0(E)$.

We focus on the retarded Green function $G^{R}_L(\omega)$ relative to the paramagnetic solution in order to calculate the spectral function $A_L(\omega)=-2 \Im{G^{R}_L(\omega)}$. The retarded Green function within the Lacroix approximation for large $U$ \cite{Lacroix,Haug} is 
\begin{eqnarray}
G^{R}_L(\omega) &=& \frac{1-\rho}{\hbar \omega - \epsilon -\Sigma_0(\hbar \omega)-\Sigma_h(\hbar \omega)}+  \nonumber \\
&& \frac{\rho}{\hbar \omega - \epsilon -U -\Sigma_0(\hbar \omega)-\Sigma_p(\hbar \omega)},
\label{functiong}
\end{eqnarray}
where $\rho$ is the level density per spin self-consistently calculated through the following integral 
\begin{equation}
\rho=\int_{-\infty}^{+\infty} \frac{d (\hbar \omega)}{2 \pi i} G^<_L(\omega),
\end{equation}
with the equilibrium lesser Green function $G^<_L(\omega)$
\begin{equation}
G^<_L(\omega)=i f(\hbar \omega) A_L(\omega),
\end{equation}
and $f(\hbar \omega)=1/(\exp{[\beta (\hbar \omega-\mu)]}+1)$ the free Fermi distribution corresponding to the average chemical potential 
$\mu=0$. In Eq. (\ref{functiong}), the self-energy $\Sigma_h(\hbar \omega)$ is
\begin{equation}
\Sigma_h(\hbar \omega)=- \frac{U \Sigma_1(\hbar \omega)}{\hbar \omega-\epsilon-U-\Sigma_0(\hbar \omega)-\Sigma_3(\hbar \omega)},
\label{sigh}
\end{equation}
while the self-energy $\Sigma_p(\hbar \omega)$ is
\begin{equation}
\Sigma_h(\hbar \omega)= \frac{U \Sigma_2(\hbar \omega)}{\hbar \omega-\epsilon-\Sigma_0(\hbar \omega)-\Sigma_3(\hbar \omega)},
\label{sigp}
\end{equation}
where the self-energy $\Sigma_i(\hbar \omega)$, with $i=1,2,3$, is given by
\begin{eqnarray}
&& \Sigma_i(\hbar \omega)=\int_{-\infty}^{+\infty} \frac{d E}{2 \pi} \Gamma_i(E) \times  \nonumber \\
&& \left[ \frac{1}{\hbar \omega +E -\mu-2 \epsilon-U+i\eta}
+\frac{1}{\hbar \omega -E +\mu+i\eta}  \right],
\end{eqnarray} 
with $\Gamma_1(E)=\Gamma_0(E) f(E)$, $\Gamma_2(E)=\Gamma_0(E) [1-f(E)]$, $\Gamma_3(E)=\Gamma_0(E)$, and 
$\eta \rightarrow 0^+$. We notice that, for large $U$, the weights of the poles of the Green function in Eq. (\ref{functiong}) are the same of the Green function examined in the main text. The Green function within the Lacroix approach has the additional self-energy terms 
$\Sigma_i(\hbar \omega)$, which take into account tunneling processes back and forth to the leads. 

\begin{figure}
\centering
\includegraphics[width=8.5cm,height=9.0cm]{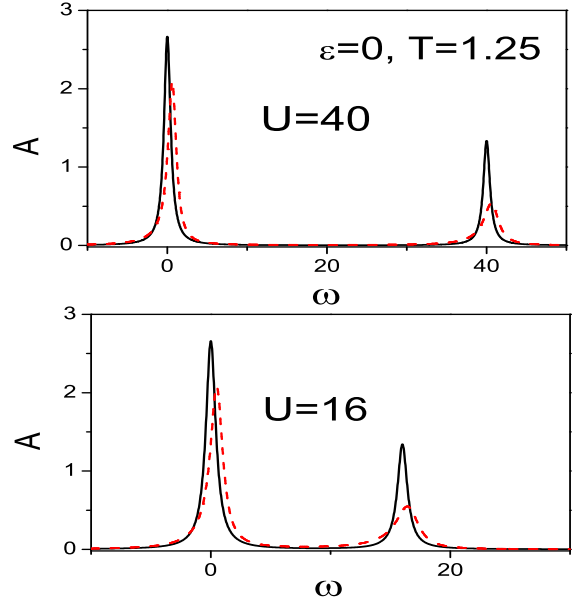}
\caption{(Color online)  Spectral function $A$ (in units of $1/ \hbar \Gamma$) as a function of frequency $\omega$ (in units of 
$\Gamma$) for Hubbard interaction $U=40 \hbar \Gamma$ (Upper panel) and $U=16 \hbar \Gamma$ (Lower panel) at level energy $\epsilon=0$ and $T=1.25 \hbar \Gamma / k_B$ (close to room temperature) in the absence of electron-vibration coupling.
Solid line: first correction upon the atomic limit (used in the main text); dashed line: additional correction upon the atomic limit (Lacroix approach).}
\label{figua1}
\end{figure}

As shown in Fig.\ref{figua1}, we compare the spectral function obtained within the approach used in the main text and $A_L$ within the Lacroix approximation \cite{Lacroix} close to room temperature for two values of $U$ ($U=40$ upper panel, $U=16$ lower panel). Both  spectral functions exhibit a bimodal structure whose peaks are separated by the energy $U$. The positions of the peaks within the two approaches are very close, while the heights of the peaks are slightly different. However, the ratio of the spectral weights of the two peaks does not significantly depend on the approach. Obviously, the modification of the isolated resonances is slightly more complicated within the Lacroix approach than that due to the self-energy $\Sigma_0(\hbar \omega)$ alone. Actually, the peaks within the Lacroix approach tend to be a little bit asymmetric. Summarizing, the differences between the two approaches are minimal supporting the use of the Green function method adopted in the present work. Finally, the small differences between the two approaches are quantitatively similar with decreasing $U$ from $40$ to $16$.

\begin{figure}
\centering
\includegraphics[width=9.5cm,height=7cm]{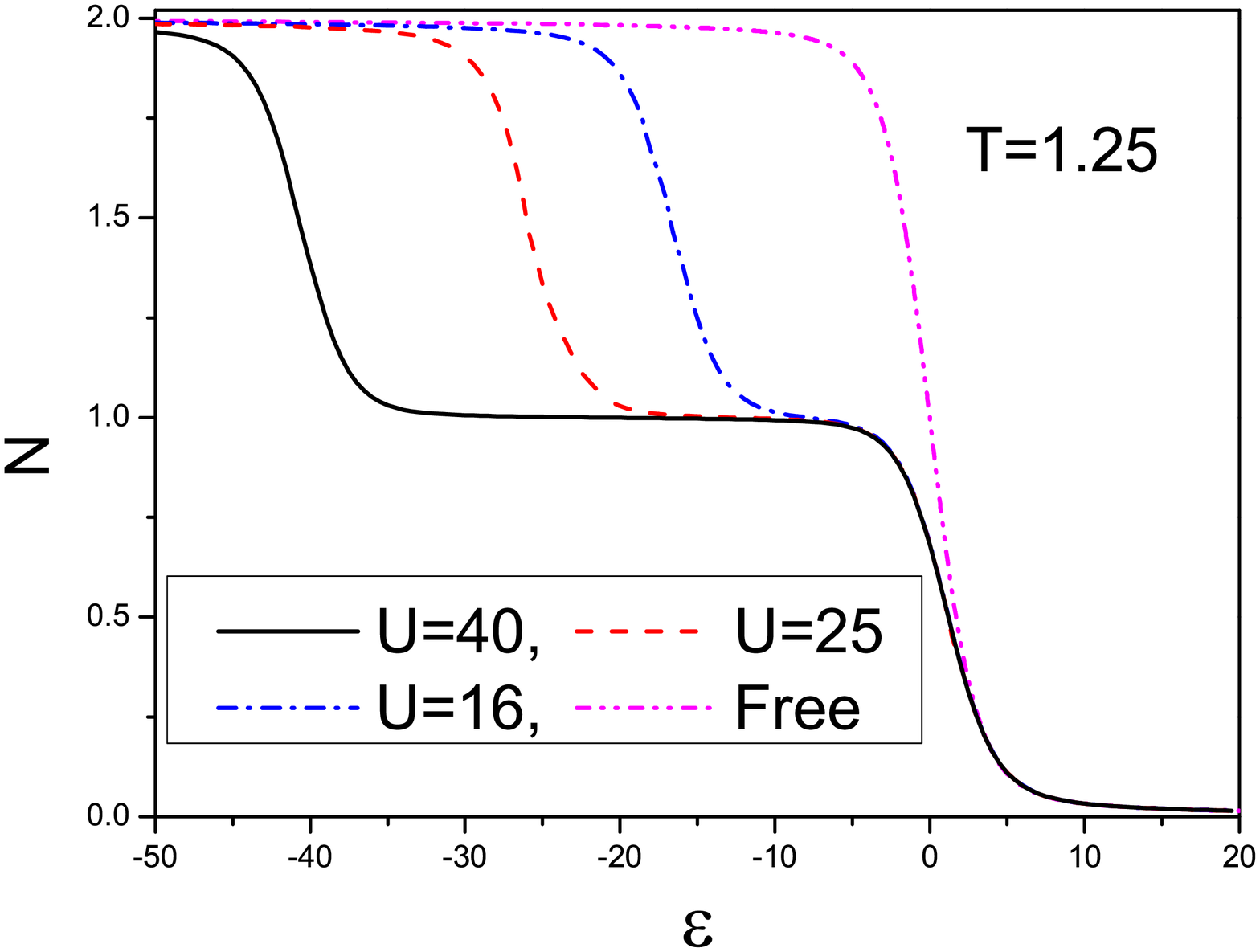}
\caption{(Color online) Level density $N$ as a function of level energy $\epsilon$ (in units of $\hbar \Gamma$) for different values of the Hubbard interaction $U$ (in units of $\hbar \Gamma$) at $T=1.25 \hbar \Gamma / k_B$ (close to room temperature) in the absence of electron-vibration coupling.}
\label{figua2}
\end{figure}

In this Appendix, we analyze also the total level occupation $N=2 \rho$ (within the paramagnetic solution). This quantity has been calculated by the two approaches discussed in this Appendix finding minimal differences. In Fig.\ref{figua2}, we report the occupation determined by the approach used in the main text as a function of level energy $\epsilon$ for different values of $U$.  It shows the typical profiles of the Coulomb blockade. Actually, for level energy $\epsilon$ around $-U/2$, $N$ is $1$. The energy region with occupation close to $1$ gets enhanced with increasing the value of $U$. Moreover, for $\epsilon$ around $\mu=0$, $N$ goes from $1$ to $0$, while,
for $\epsilon$ around $-U$, there is the transition from $N=2$ to $N=1$. These particular values of $\epsilon$ are carefully analyzed in the main-text when the effects of the electron-vibration coupling are included.

\end{appendix}


\end{document}